\documentclass[10pt, conference, compsocconf]{IEEEtran}
\usepackage{cite}

 \usepackage{graphicx}

\begin{document}

\def\b{\begin{equation}}
\def\e{\end{equation}}
\def\ba{\begin{eqnarray}}
\def\ea{\end{eqnarray}}
\def\eg{{\it e.g.~}}
\def \ie {{\em i.e.~~}}
\def \ggeq {\lower0.9ex\hbox{ $\buildrel > \over \sim$} ~}
\def \aap {AAP}
\def \apj {ApJ}
\def \mnras {MNRAS}
\def \apjs {ApJS}

%
\title{Geometry of the Cosmic Web: Minkowski Functionals from the Delaunay Tessellation}


\author{\IEEEauthorblockN{Miguel A. ~Arag\'on-Calvo}
\IEEEauthorblockA{Department of Physics and Astronomy\\
The Johns Hopkins University, \\
3701 San Martin Drive\\
Baltimore, MD 21218, USA\\
miguel@pha.jhu.edu}
\and
\IEEEauthorblockN{Sergei F. Shandarin}
\IEEEauthorblockA{Department of Astronomy\\
The University of Kansas\\
Lawrence, KS 66045, USA\\
sergei@ku.edu}
\and
\IEEEauthorblockN{Alexander  Szalay}
\IEEEauthorblockA{Department of Physics and Astronomy\\
The Johns Hopkins University, \\
3701 San Martin Drive\\
Baltimore, MD 21218, USA\\
szalay@jhu.edu}
}

%

\maketitle

\begin{abstract}
We present a novel method for computing the Minkowski Functionals from isodensity surfaces extracted directly from the
Delaunay tessellation of a point distribution. This is an important step forward compared to the previous cosmological
studies when the isodensity surface was built in the field on a uniform cubic grid and therefore having a uniform spatial
resolution. The density field representing a particular interest in cosmology is the density of galaxies which is 
obtained from the highly nonuniform distribution of the galaxy positions. Therefore, the constraints caused by the
spatially uniform grid put severe limitations on the studies of the geometry and shapes of the large-scale objects:
superclusters and voids of galaxies. Our technique potentially is able to eliminate most of these limitations.
The method is tested with some simple geometric models
and an application to the density field from an N-body simulation is shown.
\end{abstract}


%
\IEEEpeerreviewmaketitle

\section{Introduction}
 Redshift surveys
of galaxies have shown that even on the largest observed scales the bulk of matter in the
universe is concentrated in clusters and superclusters of galaxies
which are separated by  huge almost empty regions,  called voids of galaxies.  
Cosmological N-body simulations in the currently popular $\Lambda$CDM model
reveal that at moderate density thresholds  $\delta \sim 3$, the
supercluster network percolates, while occupying a small fraction ($\sim$ 3\%) of
the total volume \cite{sh-hab-hei-09}.
At greater density thresholds the structure breaks  into disconnected parts
most of which are clusters of galaxies and long filaments that  link
several clusters  of galaxies together similar to structures shown in Fig. \ref{fig:two_largest}. 
They are often referred to as superclusters of galaxies. 
At even greater thresholds only the clusters of galaxies are seen as 
islands in a sea.
 
A number of statistical measures have been suggested to quantify the
patterns made by spatial concentrations of galaxies.  
The traditional approach to quantify clustering
makes use of the hierarchy of correlation functions \cite{pee-80}.
Unfortunately they become cumbersome to evaluate beyond 
the three point function and are not particularly revealing 
as geometry and shapes are concerned.
Various statistics were proposed to quantify the geometrical and
topological properties of the structure: e.g.
percolation analysis \cite{sh-83,z-ein-sh-82},
counts in cells \cite{jan-dem-83,lap-gel-huc-91}, 
minimal spanning tree \cite{bar-son-bha-85}, 
the genus measure \cite{got-mel-dic-86}, 
wavelet analysis \cite{kandu}, Multiscale Morphology Filters
from the Hessian of the density field \cite{Aragon07} and potential field \cite{Hahn07a},
and recently the SpineWeb algorithm \cite{Aragon08} and the 
skeleton of the cosmic web \cite{nov-etal-06, sou-etal-08}, etc.

Mecke, Buchert and Wagner (1994) have introduced 
the Minkowski functionals (hereafter MF) to cosmology. 
The four Minkowski functionals computed for the excursion sets at many
density thresholds contain valuable information regarding both the
geometrical as well as topological distribution of matter in the Universe.
There have been three major attempts made to study the morphology of
the Cosmic Web using MF; these efforts differ in their approach of evaluating
the MF: (1) Firstly, Boolean grain models study the MF of surfaces which result due
to intersecting spheres decorating the input point-set \cite{mec-buc-wag-94}. 
(2) Secondly, Krofton's formulae make it possible to calculate MF on a
density field defined on a grid \cite{sch-buc-97}. In
this case the MF are calculated by using the information of the number
of vertices, edges,  faces and cuboids. (3) Finally, an
alternative, resolution-dependent approach consists in employing the
Koenderink invariants \cite{sch-buc-97}.

A radically new and powerful approach for using the Minkowski functionals 
in cosmology was firstly introduced for the analysis of two-dimensional CMB 
(Cosmic Microwave Background) maps \cite{nov-fel-sh-99, sh-etal-02}.
This approach consists of constructing isotemperature  contours and analysis 
of individual parts of the excursion set at many levels. These
contours are triangulated and the MF are evaluated for the resulting
closed polygons. Since Minkowski functionals are additive in nature one can
glean information regarding both individual objects (galaxies,
clusters, voids) as well as the supercluster-void network in its
totality.

However,  a three-dimensional 
algorithm required a new component that builds a surface of approximately 
constant (to linear order) density. It was implemented in 
\cite{she-etal-03} and then used for the analysis of the mass distribution
at the non-linear stage in cosmological N-body simulation of the $\Lambda$CDM model
\cite{sh-she-sah-04} and mock  galaxy catalogs \cite{she-mn-04, she-phd-04}.
In addition, particular ratios of Minkowski
functionals quantifies the {\em morphology} of large scale structure
by telling us whether the distribution of matter in
superclusters/voids is spherical, planar, filamentary etc. \cite{sah-sat-sh-98}.
The  numerical technique used in the cosmological studies is based on the marching cubes 
algorithm (MCA) in three-dimensional density field specified on a uniform mesh.
It therefore has a uniform resolution scale over the region in question.
The data points are  the positions of the galaxies which are distributed in highly  
non-uniform manner:
there are clusters of galaxies where thousands of galaxies packed 
in the volume of a few Mpc$^3$ and there are voids of galaxies where 
hardly one galaxy can be found in hundreds of Mpc$^3$.
Thus, for the study of the densest clusters of galaxies one needs to have
a fine mesh, but it does not work in voids were most of sites are empty.
On the other hand a mesh with large cells would allow to study the voids
but erase the clusters. A numerical method based on adaptive mesh
is badly needed for the studies of the structure in the universe.

\subsection{Extracting implicit surfaces from scalar fields}

The extraction of implicit surfaces from a given scalar field is an important
problem for visualization and volumetric data analysis. The marching cubes algorithm
\cite{Lorensen87} was introduced in order to extract isosurfaces from
a regular grid by identifying intersections between each voxel of the grid and
an isosurface defined by a threshold value $\delta_t$
and then generating a triangular mesh from the intersecting points. 
The MCA is an efficient and fast solution but it presents a few ambiguities. 
In addition to this its use was until recently restricted by a patent. In order to 
circunvent the patent restrictions and solve the ambiguities in the MCA 
the Marching Tetrahedra Algorithm (MTA) was developed \cite{Doi91, Gueziec95}.
The main idea behind MTA is to divide each voxel of the regular grid into 6 
tetrahedra and identify the intersections between each tetrahedron and the
isosurface defined by $\delta_t$. This greatly simplifies the identification of intersections since there are
only 8 cases to consider: no intersection, 4 intersections with a triangle and
3 intersections with two planar triangles or quad (see Fig. \ref{fig:tetra_iso_itersection}). 
There are no ambiguities in the intersection cases and they can be conveniently encoded in a lookup table.
Both MCA and MTA use linear interpolation in order to identify the intersecting 
points.

The MCA and MTA take as input a scalar field sampled on a regular grid. 
In many of the applications of MCA and MTA the scalar
field is computed directly in a previous step, like in tomography and MRI medical
images or analytic functions for 3D texture and surface generation in computer images. 
In the astronomical context often the scalar field (density) must be reconstructed from 
a point distribution such as mass particles from N-body simulations or 
galaxy positions from redshift surveys. The accurate reconstruction of such datasets is not
trivial an in general one has to rely on a series of assumptions about the
underlying scalar field such as continuity and linearity.

A major improvement in the reconstruction and analysis of density fields from point distributions came with the implementation of 
Voronoi and Delaunay tessellation techniques \cite{Voronoi1908,Delaunay34,Okabe00, Bremer04, Edels03}. In particular for astronomy the
Delaunay Tessellation Field Estimator (DTFE) \cite{Schaap00, Schaap07,Weygaert09} 
based on the previous work of \cite{Bernardeau96}  which applied the Delaunay tessellation for
the interpolation of a field sampled by an (in principle) arbitrary point distribution.
The DTFE estimates local densities at each point as being inversely proportional to
the volume of the adjacent Voronoi cell of the point. The density field is then
sampled on a regular grid by linearly interpolating the density 
field inside each tetrahedron containing the sampling point. 
The DTFE represents a natural alternative to grid-based methods as it adapts itself to the point 
distribution, it produces a space-filling density field and is well defined for all regions inside 
the volume containing the points.
However, the DTFE is also known to produce strong artifacts as a result of the linear interpolation of the density field
over very large and thin triangles. By construction it can not provide zero density estimates, therefore it can not reproduce 
the density field inside very underdense regions where there are no sampling points. 
In its original implementation it samples the density field on a regular grid thus
the final density field is not adaptive and due to the discrete sampling scheme it introduces
aliasing in the density field.

The use of a regular grid is not required in the case when we already have
values of the scalar field at each sampling point. By recognizing that the
Delaunay interpolation schema does not require the coupling with the point
distribution (as in the DTFE) one can use the tessellation to interpolate any scalar field 
independently of the sampling points. Here we use a Delaunay tessellation to describe the
connectivity between points that will be used to interpolate the density
and extract isosurfaces. The density field can be computed
by adaptive methods such as the scale-adaptive DTFE, Spline or fixed-scale Gaussian kernel.
The density values at each vertex of the tessellation are then used to identify intersections with the isosurface. 
By computing the isosurfaces directly from the Delaunay tessellation we avoid two extra steps:
i).- interpolation of the density field on a regular grid and ii).- tetrahedrisation of the voxels in the grid. 
Our algorithm can be directly applied to the Delaunay tessellation with no further processing.

\section{Isosurfaces from the Delaunay Tessellation: Marching tetrahedra Algorithm}
\label{sec:march_tetra}

The first step in the reconstruction of isosurfaces is the computation of the
Delaunay tessellation of the point distribution which will
be used to define the spatial connectivity between points used to interpolate the density field.
The tessellation is also used to generate adaptive density estimates at each point as
described in \cite{Schaap00}

\subsection{Isosurface extraction}

\begin{figure}[!fbt]
  \centering
  \includegraphics[width=0.5\textwidth,angle=0.0]{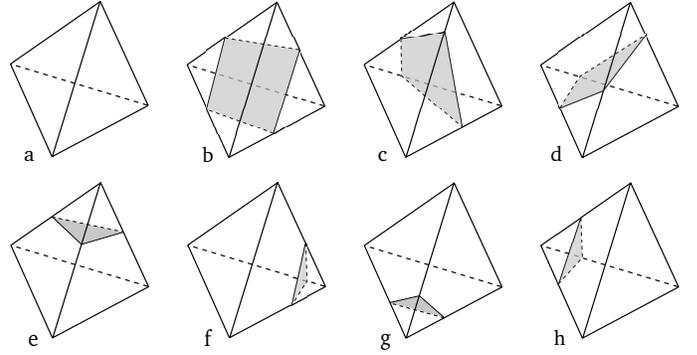}
    \caption{Possible intersection cases between a tetrahedron and the isosurface. 
      \textbf{a}: no intersection with tetrahedron, \textbf{b}-\textbf{d}: intersection in a quad, there are two 
      more dense vertices and two less dense.
      \textbf{e}-\textbf{h}: intersection in a triangle, there is one single denser or less dense vertex.}
  \label{fig:tetra_iso_itersection}
\end{figure}

Once we computed the Delaunay tessellation we test each tetrahedron for intersection with the 
isosurface defined by the threshold value $\delta_t$.
We do this by simply evaluating $\textrm{min}(\delta_i) \le \delta_t \le \textrm{max}(\delta_i)$, 
where $\delta_{i=0...3}$ is the density field evaluated at each of the four vertices of the tetrahedron. We then proceed to 
count the number of edge intersections. There are 7 intersecting cases (see Fig. \ref{fig:tetra_iso_itersection}), 
when the isovalue divides the tetrahedron in two lower values and two higher values the tetrahedron is cut by a quad, 
otherwise it is cut by a single triangle. Each intersecting point corresponds to a vertex of the triangulated isosurface.
The position of the intersecting point is estimated by interpolating the field between the two 
vertices ($\mathbf{x_0}$, $\mathbf{x_1}$) of the intersected edge as follows:

\begin{equation}
 \begin{array}{rcl}
  x &=& (1-t) \; x_0 + t \; x_1\\
  y &=& (1-t) \; y_0 + t \; y_1\\
  z &=& (1-t) \; z_0 + t \; z_1
 \end{array}
\end{equation}

\noindent where the parameter $t$ runs in the interval $[0,1]$ and is given by:

\begin{equation}
  t = (\delta_t - \delta_0) / (\delta_1-\delta_0).
\end{equation}

\noindent Depending on the number of intersecting points we construct a triangle (3 intersections) or pair of coplanar triangles for a quad (4 intersections). 
The vertices of the triangle are then oriented to have their normal vector pointing outside the surface i.e. 
in the direction of decreasing density. This can be easily checked with the inner product between the normal
of the triangle and the less dense vertex of the tetrahedron. The orientation of the quads is also checked to 
avoid incorrect triangulations when the vertices are not properly oriented as adjacent triangles.

\section{Implementation}

The Delaunay tessellation of the point distribution was computed using the publicly 
available CGAL (www.cgal.org) library. This efficient library allows us to compute the tessellation for
$256^3$ particles in a couple of minutes on a regular linux workstation. Our implementation can generate 
open or periodic boundary conditions by using a buffer region around the original box 
and replicating the indexes of each point. Note that the most recent CGAL release includes periodic boundaries natively.
This new feature will be included in the next version of our code. For the present application our approach 
is sufficient.

The code that constructs the isosurface from the Delaunay Tesselllation was written as
part of a larger visualization suite written in C++ built on top of the OpenFrameworks 
(http://www.openframeworks.cc) environment for OpenGL-based volume and surface renderings 
as well as multimedia mixing for data sonification, sound visualization, etc. 
(see Fig. \ref{fig:models_all}). 

A version of the MTA was also written in the IDL language for fast development and testing of the algorithms.
The code loops over all tetrahedra and constructs triangles or quads as described in 
section \ref{sec:march_tetra}. This raw surface consists of unconnected  triangles so we
proceed to remove duplicated vertices by indexing them. Repeated vertices are identified by
means of the edge from which they were created. Al vertices sharing the same parent edge 
correspond to the same point.

\subsection{Isolated Object Extraction}
\label{sec:object_label}

In general, the isosurface consists of a set of isolated objects. In order to study the properties of
each individual object we must be able to identify it. This is done by first constructing a list of adjacent triangles. 
For each triangle in the isosurface we iteratively link and label its three adjacent triangles until all 
triangles are connected. The final triangulation divided into isolated objects is stored in a binary format 
and as a Waveform .obj format file that can be used for analysis and visualization. 

\subsection{Density estimation}

In order to interpolate the density field inside each tetrahedron we need density estimates at each point.
There are several methods for this purpose, among them the n-nearest neighbor, top-hat counts, Gaussian
kernel and DTFE are the most common \cite{Bernardeau96,Sibson81,Sukumar98}. Adaptive density estimators such as the 
DTFE are strongly dependent on the spatial sampling. This is more evident in low density regions where one can see strong artifacts
arising from the linear interpolation inside the tetrahedron. The use of averaging procedures on the tessellation or
higher order interpolation alleviates some artifacts at the expense of increased complexity of the method.
The end result is that isosurfaces extracted from adaptive density estimates tend to be highly irregular in
low density regions. 
For the analysis presented in the coming sections we choose a fix-scale Gaussian kernel density estimator. 
This limits out ability to probe into compact high density regions such as the interior of clusters. 
However, for low-intermediate density regions the Gaussian smoothing produces better surfaces than
the adaptive approaches like the DTFE provided that the smoothing kernel is larger than the mean interparticle separation.
By using a Gaussian kernel we still take advantage of the space-filling properties of the tessellation
and the convenient use of tetrahedra to interpolate the density field and identify intersections with the 
isosurface. In a following paper we will present a more detailed comparison between adaptive and fixed-scale
density estimation. For the purposes of this work we restrict our analysis to Gaussian densities.

We compute Gaussian densities at particle $i$ as the sum of a Gaussian kernel evaluated at each of the surrounding 
$j$ particles:

\begin{equation}
  \rho_i = \sum_j e^{-(x_j-x_i)^2 / (2 \; \sigma^2)}
\end{equation}

\noindent where $\sigma$ is scale of the smoothing kernel. For convenience we use the overdensity defined as:

\begin{equation}
\delta = \frac{\rho_i }{\hat{\rho}}.
\end{equation}

\noindent Where $\hat{\rho}$ is the mean density computed for the given $\sigma$.

The density estimation is an $O(N^2)$ operation, which makes the brute-force implementation unfeasible 
for large datasets. We use a bucket algorithm where the containing box is divided into a coarse grid. Instead of evaluating 
the Gaussian kernel for all the particles we restrict the operation to particles inside bucket containing particle $i$
as well as its 26 adjacent buckets. We use a  grid size such that there is a warranty of enclosing particles up to $5\sigma$ away from the 
target particle. Our fully threaded code greatly improves the speed of the density estimation and scales almost 
linearly with the number of processors.

The isodensity surfaces computed assuming a linear variation of the density field inside each tetrahedron tend to be
very irregular. This effect is even more pronounced in the interface regions between high and low density cosmological structures such
as the edges of walls, filaments and clusters where the density field profile is close to a power law. 
In these regions, if the spatial sampling is low, there can be a large variation in 
the density field inside an individual tetrahedron. Also, the overall density distribution
of the evolved Cosmic Web is roughly log-normal \cite{Coles91}.
As a result of this, the isodensity surfaces computed from linear interpolation tend to be highly irregular 
and to overestimate the enclosed volume inside the surface. Based on the character of the density distribution and the power-law behavior
of the density profiles it makes more sense to use a transformation that linearizes the field 
inside the tetrahedron. We therefore use the $log(\delta)$ and linearly interpolate the log-transformed density field. 
The $log(\delta)$ interpolation does not conserve mass as in the linear case but 
it produces smoother surfaces and better volume estimation.

\section{Morphological Parameters}
\subsection{Minkowski Functionals}
\label{sec:morph_param}

In this work we discuss the geometry and topology of the regions bounded
by the isodensity surfaces and therefore make no prior assumptions about
the shapes of superclusters and voids.  It is worth noting that some
regions may have more than one boundary surface and possess nontrivial
topology of the boundaries.  The complete characterization of an
arbitrarily complex region in three dimensions obviously cannot be
achieved if only a few numbers are used. At best one can try to design
some basic characteristics that serve a particular purpose. Our
purpose is to provide basic measures suitable for quantification of
typical components of the large-scale structure: superclusters and voids.

Four Minkowski functionals are effective non-parametric descriptors of
the morphological properties of surfaces in three dimensions
\cite{mec-buc-wag-94,mat-03,she-etal-03}. They are
\begin{itemize}
\item {\it Volume} $V$ enclosed by the surface $S$,
\item {\it Area} $A$ of the surface,
\item {\it Integrated mean curvature} $C$ of the surface, 
\b
\label{eq:curv}
C = \frac{1}{2}\oint_S{\left({1\over R_1} + {1\over R_2}\right)da}, 
\e
where $R_1$ and $R_2$ are the principal radii of curvature at a given
point on the surface.
\item {\it the Euler characteristic}
\b
\label{eq:euler} \chi = \frac{1}{2\pi}\oint_S{\left({1\over
      R_1R_2}\right)da}.  
\e
\end{itemize}

\subsection{Shapefinders}

As demonstrated in \cite{sah-sat-sh-98,sat-sah-sh-98} particular ratios of Minkowski
functionals called shapefinders provide us with a set of
non-parametric measures of sizes and shapes of objects.  Therefore, in
addition to determining MFs we shall also derive the shapefinders, $T$
(Thickness), $B$ (Breadth) and $L$ (Length) defined as follows:
\b \label{eq:TBL}
T = {3V\over A}, \ \  B = {A\over C}, \ \ L = {C\over 4\pi}.
\e 
This normalization gives roughly a half of the diameter in the case  of anisotropic objects. 
The three shapefinders describing an individual region bounded by one
or several isolated surfaces of constant density have dimensions of
length and provide us with an estimate albeit quite crude
of the region's `extensions': $T$ is the
shortest and thus describes the characteristic thickness of the region or
object, $L$ is typically the longest and characterizes the length of
the object; $B$ is intermediate and can be associated with the breadth
of the object.  This simple interpretation is obviously relevant only
for fairly simple shapes.  The choice of the coefficients in 
(\ref{eq:TBL}) results in a sphere having all three sizes equal to its
radius $T=B=L=R$. A triaxial ellipsoid has values of $T$, $B$ and $L$
close but not equal to the lengths of its three principal semi-axes:
shortest, intermediate and the longest respectively.  It is worth
noting that $T$, $B$ and $L$ are only the estimates of three basic
sizes (semi-axes) of an object which work quite well on such objects
as a triaxial ellipsoid and torus \cite{sah-sat-sh-98,sat-sah-sh-98,she-etal-03}
but no three numbers can describe an arbitrary, complex
three-dimensional shape. In particular, the length of a region with many
tunnels may be better characterized by $\tilde{L}=L/(1-\chi/2))$, because
$(1- \chi/2)$ is equal to the number of tunnels and therefore $\tilde{L}$ 
approximately represents the length of the parts between tunnels.

As suggested in \cite{sah-sat-sh-98} a rough idea about the global `shape' 
of a region can be provided by an additional pair of dimensionless shapefinders:
\b 
\label{eq:shapefinder} 
P = {B-T\over B+T};~~~ F = {L-B\over L+B}, 
\e
where $P$ and $F$ are measures of Planarity and Filamentarity
respectively ($P, F \leq 1$).  A sphere has $P = F = 0$, an ideal
filament has $P = 0, F = 1$ while $P = 1, F = 0$ for an ideal pancake.
Other interesting shapes include `ribbons' for which $P \sim F \sim
1$.  When combined with the genus measure, the triplet $\lbrace P,F,G
\rbrace$ provides an example of {\em shape-space} which incorporates
information about topology as well as morphology of superclusters and voids. 

Note however that non-geometrical shapefinding statistics based on mass moments
etc. can give misleading results when applied to large scale
structure, as demonstrated in \cite{sat-sah-sh-98}.

\subsection{Numerical Estimates of Minkowski Functionals}
We construct a 2-polytopal surface using an assembly of triangles in
which every triangle shares its sides with each of its three neighboring
triangles. Here we outline the technique used for the numerical estimate
of the Minkowski Functionals. It was suggested in \cite{she-etal-03},
where it is described in more detail, see also \cite{she-phd-04}
\begin{itemize}
\item The area of such a triangulated surface is
\b 
A = \sum_{i=1}^{N_F}{A_i}, 
\e
where $A_i$ is the area of the $i^{th}$ triangle face and $N_F$ is the
total number of faces which compose a given surface.
\item The volume enclosed
by this 2-polytopal surface is the summed contribution from $N_F$
tetrahedra
\ba \label{eq:volume}
V &=& \sum_{i=1}^{N_F}{V_i},\nonumber\\
V_i &=& \frac{1}{3} A_i (n_j{\bar P}^j)_i.
\ea
Here $V_i$ is the volume of an individual tetrahedron whose base is
a triangle on the surface.
$(n_j{\bar P}^j)$ is the scalar product between the outward 
pointing normal $\hat{n}$ to this triangle and the mean position 
vector of the three triangle vertices, for which the $j^{th}$
component is given by
\b
\label{eq:volume1}
{\bf{\bar P^j}} = \frac{1}{3}(P_1^j + P_2^j + P_3^j).
\e
The subscript $i$ in (\ref{eq:volume}) refers to the $i-$th
tetrahedron, while the vectors $P_1, P_2, P_3$ in (\ref{eq:volume1})
define the location of each of three triangle vertices defining the
base of the tetrahedron relative to an origin. The origin can be
arbitrary chosen.

\begin{figure*}[!hfbt]
  \centering
  \includegraphics[width=0.99\textwidth,angle=0.0]{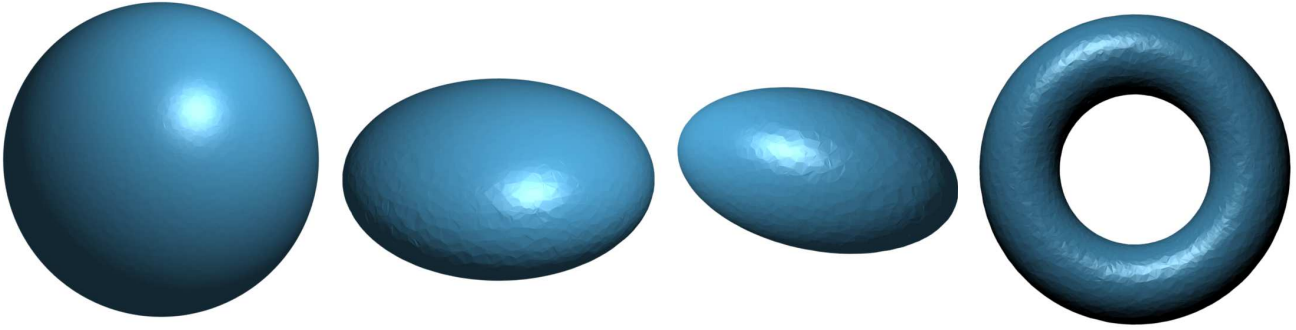}
    \caption{Four simple geometries used for testing, from left to right: sphere, pancake, filament and torus.}
  \label{fig:models_all}
\end{figure*}

\item The extrinsic curvature of a triangulated surface is localized in the
triangle edges. As a result the integrated mean curvature $C$ is
determined by the formula
\b \label{eq:curvature}
C = {1\over2}\sum_{i,j}{\ell_{ij} \cdot \phi_{ij}} \cdot \epsilon, 
\e 
where $\ell_{ij}$ is the length of the edge common to triangles $i$ and $j$ and 
$\phi_{ij}$ is the angle between the normals $\hat{n}_i$ \& 
$\hat{n}_j$ to these adjacent triangles
\b 
\cos\phi_{ij} = \hat{n_i} \cdot \hat{n_j}.  
\e
The summation in (\ref{eq:curvature}) is carried out over all edges. 
It should be noted that for a completely general
surface, the extrinsic curvature can be positive at some (convex)
points and negative at some other (concave) points on the surface. 

\item The Euler characteristics of a  polyhedral surface is given by 
\b
\label{eq:EC}
\chi = N_F - N_E + N_V, 
\e
where $N_F, N_E, N_V$ are, respectively, the total number of faces, 
edges, and vertices defining the surface. Formula \ref{eq:EC}
is correct for arbitrary shaped faces, obviously it can be reduced to 
\b
\label{eq:EC-triangle}
\chi = N_V - N_F/2, 
\e
in the case when all faces are triangles since the number of edges is
$N_E = 3N_F/2$, which also holds true for polytopal surfaces.
\end{itemize}

\section{Results}

In this section we present two set of tests focused on simple geometries and a more complex
surface extracted from a cosmological N-body simulation.

\subsection{Simple Models}

We constructed a set of simple geometries in order to test our method. Three geometries
represent the generic types of the gravitational collapse in the Zel'dovich 
framework \cite{z-70,sh-z-89}:
a compact clump represented by a sphere, pancake and filament. Additionally we construct a torus in order to test its topology.
The geometries were constructed by generating a set of hypothetical sampling points
over a continuous scalar field defined as:

\begin{equation}
d = \left( {x^2 \over a^2} + {y^2 \over b^2} + {z^2\over c^2} \right)^{1/2}.
\end{equation}

\noindent Where $d$ is the scalar value of the field and we define the following limiting cases:

\begin{equation}
\begin{array}{lcl}
  a=b=c & : &\textrm{ sphere }\\
  a\approx b \gg c & : &\textrm{ pancake }\\
  a\gg b \approx c & : &\textrm{ filament.}
\end{array}
\end{equation}

\noindent The torus was computed as:

\begin{equation}
d = \left( \left(R - \left(x^2 + y^2 \right)^{1/2} \right)^2 + z^2 \right)^{1/2}.
\end{equation}

\noindent Where $R$ is the radius of the torus and $t$ is the radius of the tube.
We evaluated the distance function for each of the sampling points and proceed to
construct the isosurface at a defined threshold value. Since the isosurfaces defined on the distance field
enclose underdense regions we flip the normals so that they point outside the surface.
Note that in this test surfaces the sampling points are not coupled with the scalar field.
They are used only to define the tessellation that will be used to interpolate the density field.

Fig. \ref{fig:models_all} shows the four models after we extraced the isosurface. We rendered the
raw surface (i.e. no normal smoothing) in order to emphasize the triangles that form the surface. 
The MTA produces very smooth surfaces although in this case it only reflects the spatial sampling
of the random points.

\begin{figure*}[!fbt]
  \centering
  \includegraphics[width=0.49\textwidth,angle=0.0]{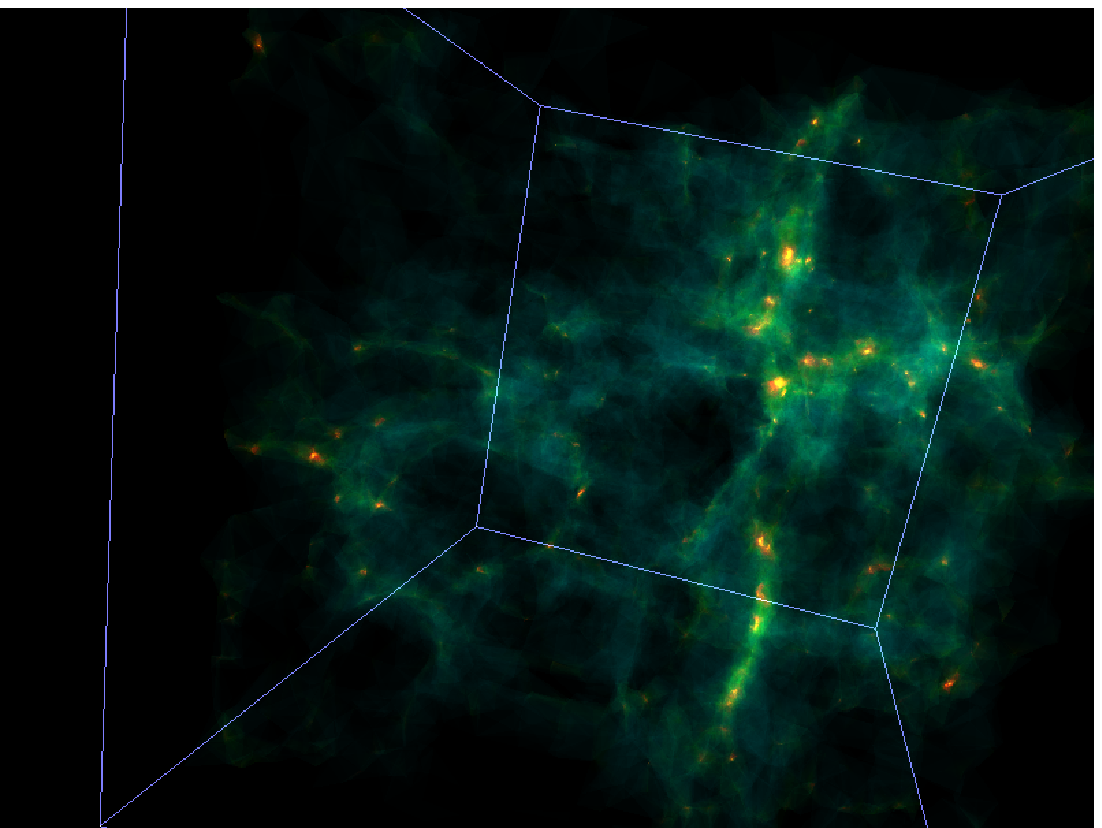}
  \includegraphics[width=0.49\textwidth,angle=0.0]{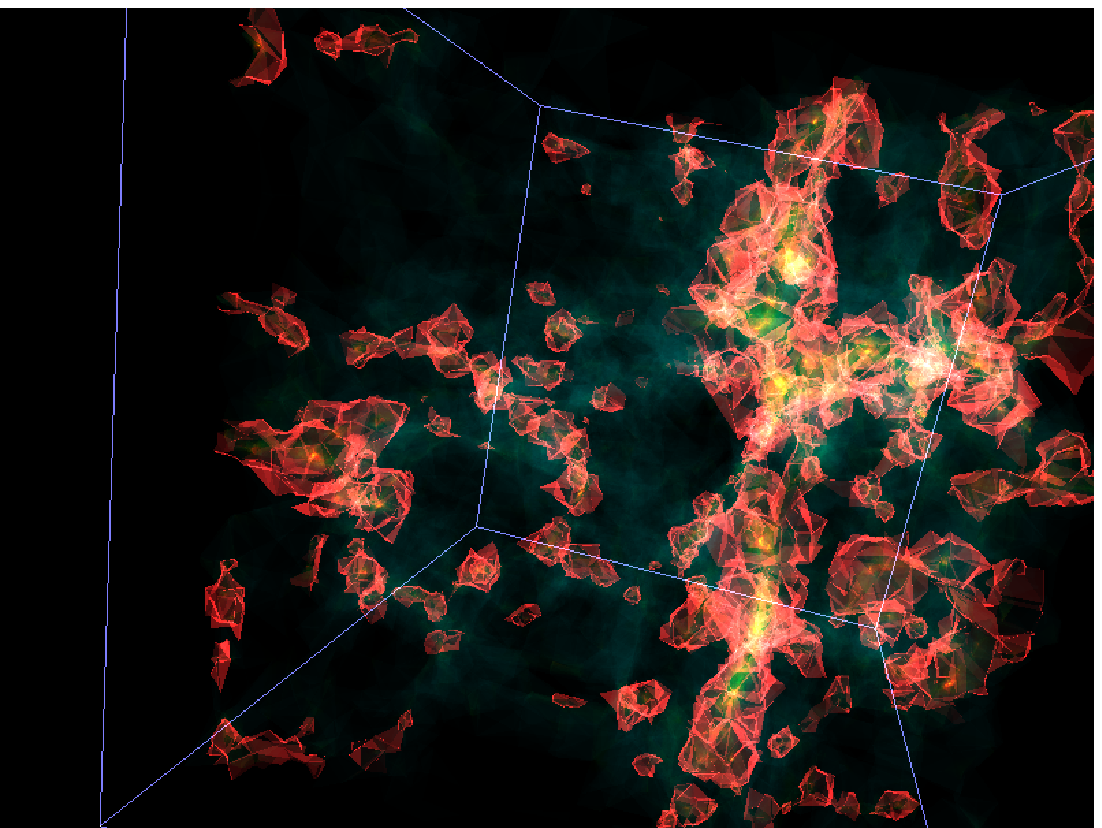}
    \caption{Volume rendering of a density field (left) and isodensity surface 
      at $\delta = 3$ (right).}
  \label{fig:volume_rendering}
\end{figure*} 

Table \ref{tab:mikowsky_and_shape} sumarizes the result of our analysis over the four geometries. 
The parameters for which there are analytical formulae (\eg all parameters  for the sphere) were computed with the accuracy better than 1\%. The numbers of faces, edges and vertices  
in the case of the sphere were 8,553, 25,653 and 17,102. The numbers of the triangulation 
parameters in the cases of the filament, pancake and torus were similar while two largest superclusters 
in the cosmological N-body simulation they were about and order of magnitude greater.

\begin{table}[!h]
  \caption{Minkowski functionals and shape measures}
  \label{tab:mikowsky_and_shape}
  \begin{center}
    \leavevmode
    \begin{tabular}{lllll} \hline \hline
      Mink Func     & Sphere  &   Pancake &   Filament & Torus\\
      Volume        & 0.065   &   0.016   &   0.004 &  0.059\\
      Area          & 0.788   &   0.449   &   0.163  &  1.183\\
      C           & 3.156   &   2.564   &   1.826  &  5.923\\
      EC            & 2       &   2       &   2  &  0 \\
      \hline        
      Thickness  & 0.248   &   0.109   &   0.075  &  0.149\\        
      Breadth       & 0.249   &   0.175   &   0.089  &  0.199\\
      Lenght        & 0.251   &   0.204   &   0.145  &  0.471\\
      \hline        
      Planarity     & 0.002   &   0.233   &   0.086  &  0.143\\
      Filamentarity & 0.002   &   0.075   &   0.238  & 0.404\\
      \hline        
    \end{tabular}
  \end{center}
   \end{table}

\subsection{Cosmological Density Field}

We applied our method to a more realistic and complex dataset consistinng of a  dark-matter N-body 
simulation of a standard $\Lambda$CDM cosmology. The simulation is contained inside a periodic box of 200 $h^{-1}$ Mpc.
Initial conditions were generated on a $256^3$ grid with 
$\Omega_m = 0.3$, $\Omega_{\Lambda} = 0.7$, $\sigma_8 = 0.9$ and $H = 73$ km s$^{-1}$ Mpc$^{-1}$.
\footnote{One megaparsec (Mpc) corresponds to $3.26\times10^6$ light years, $\Omega_m$ and $\Omega_{\Lambda}$ 
correspond to the ratio of matter and dark energy relatively to a close universe with $\Omega_{total} = 1$. The Hubble constant
$H$ measures the expansion of the Universe as a recession velocity per Mpc.}
After having set up 
the initial conditions, we follow the subsequent gravitational evolution to the present time using 
the public N-body code Gadget2 \cite{Springel05}. 
The simulation box is large enough to contain the full range of environments in the Cosmic Web, from large
underdense voids to massive dense clusters.

\begin{table}[!h]
  \caption{Minkowski functionals and shape measures of two largest superclusters in the N-body simulation}
  \label{tab:n-body}
  \begin{center}
    \leavevmode
    \begin{tabular}{lllll} \hline \hline              
      Mink Func     & Green  &   Red \\
      \hline
      Volume        & 0.00359   &   0.00202 \\
      Area          & 0.429   &   0.234   \\
      C  & 13.34   &   6.54   \\
      EC            & -10       & -6     \\
      \hline        
      Thickness & 0.0251  &0.0258   \\        
      Breadth       & 0.0322   &   0.0358   \\
      Lenght        & 1.06   & 0.521   \\
      \hline        
      Planarity     & 0.124   &   0.162   \\
      Filamentarity & 0.941   &   0.871  \\
      \hline        
    \end{tabular}
  \end{center}
   
\end{table}

From the final particle distribution we estimate the local density using a Gaussian kernel of 2 Mpc/$h$.
We then compute the Delaunay tessellation with periodic conditions and applied the MTA (section \ref{sec:march_tetra}) in order to
construct an isodensity surface corresponding to an overdensity threshold of $\delta=3$.
From the raw isosurface we label individual objects as described in section \ref{sec:object_label}
and compute their MF and shapefinders (see section \ref{sec:morph_param}).
Fig. \ref{fig:volume_rendering} shows the volume rendering of the density field inside the simulation box (left panel)
and the isosurface extracted at $\delta=3$ (right panel). The isosurface is formed by a large number of unconnected
structures enclosing overdense regions and clumps of matter.

\begin{figure*}[!fbt]
  \centering
  \includegraphics[width=0.99\textwidth,angle=0.0]{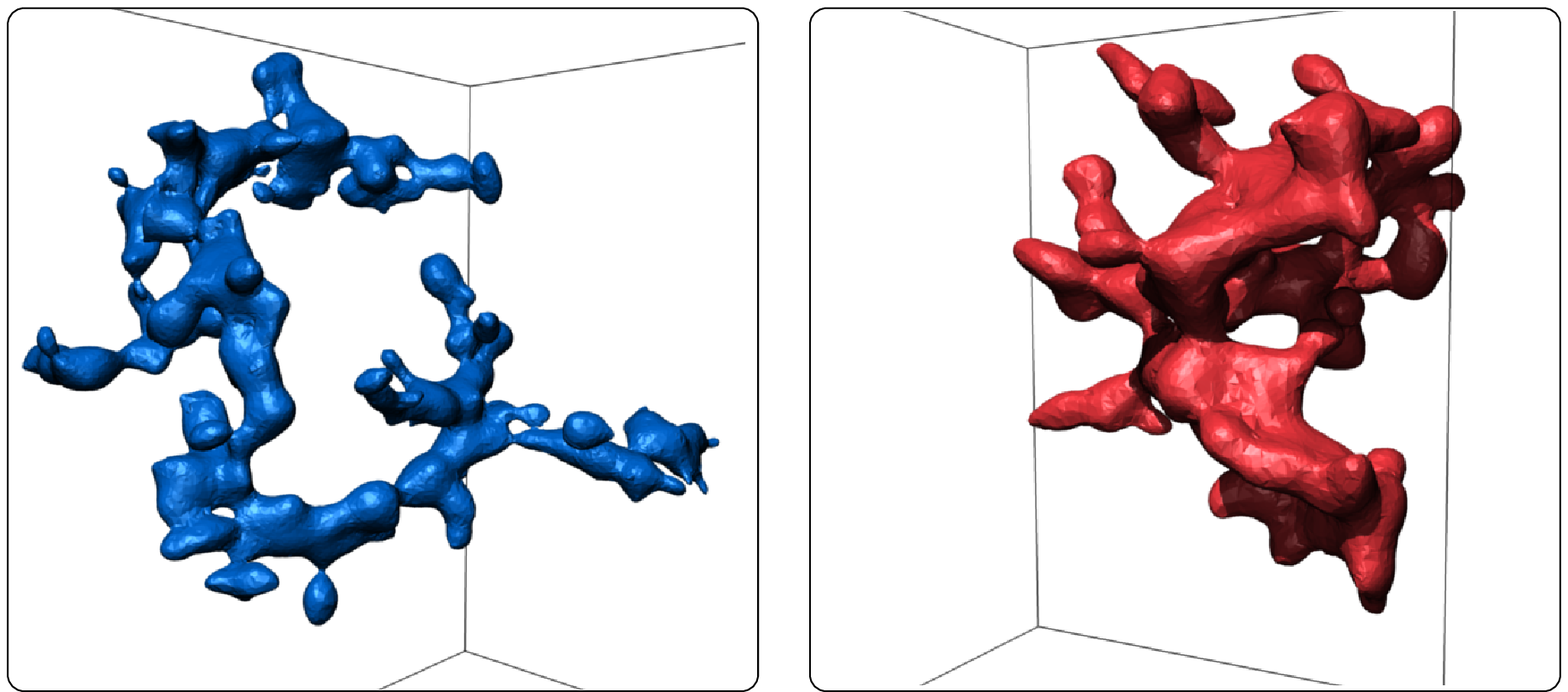}
    \caption{Largest (blue) and second largest (red) structures at $\delta_t = 3$ and Gaussian smoothing of 2 Mpc$/h$.}
  \label{fig:two_largest}
\end{figure*}

Fig.  \ref{fig:two_largest} shows the two largest superclusters depicted from the N-body simulation.
The largest structure is shown in blue and the second largest in red. 
Their parameters, MF, thickness, breadth, length and shapefinders
are given in Table \ref{tab:n-body}. Both have comparable thicknesses and breadths while the length of the 
first largest supercluster about twice longer than that of the second largest cluster. The superclusters
have 6  and 4 tunnels respectively, a couple of which can be seen in the figure. 
Although the figure gives some idea about the complexity of the objects,
its full appreciation is possible only with the help of visualization software 
where the three-dimensional structure of each can be fully recognized.

\section{Conclusion and future work}
\label{sec:conclusions}
We present a novel technique for computing Minkowski functional \ie the volume, area, integrated mean curvature and
Euler characteristic of complex objects (superclusters and voids of galaxies) that form due to the nonlinear 
gravitational growth of initial Gaussian random fluctuations in dark matter density field. These structures can be
reliably modeled in cosmological N-body simulations and are observed in large galaxy redshift catalogs like
SDSS and 2dF. The technique can also be used for other applications when quantitative information is needed
for arbitrary shaped structures. An important step forward has been made by replacing the Marching Cube 
Algorithm in previous studies \cite{she-etal-03,sh-she-sah-04,she-mn-04} by Marching Tetrahedra Algorithm for building the isodensity surface. The MCA  operates on the
fields specified on a uniform cubic grid and therefore has a uniform spatial resolution. The MTA finds the density
field directly from the positions of galaxies or particles in cosmological N-body simulations by means of
Delaunay tessellation and therefore has an adaptive spatial resolution. 
Potentially  the galaxy density field can be analyzed  with
better resolution in the regions where more galaxies are observed or more simulation particles are clustered however this requires additional testing which under way and will be 
reported in the following papers.

In this work we report the results of the basic tests that shows that the code works and ready for the studies of
the density fields obtained in cosmological N-body simulations of the structure in the universe or real structures in galaxy redshift catalogs.


\begin{thebibliography}{40}
\bibitem[1]{Aragon07}
	Arag{\'o}n-Calvo M.~A., Jones B.~J.~T., van de Weygaert R., van der Hulst J.~M.\ 2007, \aap, 474, 315
\bibitem[2]{Aragon08}
	Arag{\'o}n-Calvo M.~A., Platen, E.., van de Weygaert R., Szalay, A., arXiv:0809.5104. Unpublished.
\bibitem[3]{Bardeen86} 
	Bardeen J.~M., Bond J.~R., Kaiser N., Szalay A.~S.\ 1986, The Astrophysical Journal, 304, 15 
\bibitem[4]{bar-son-bha-85} 
	{Barrow},  J.D., {Sonoda}, D.H. \& {Bhavsar}, S.P., 1985, Monthly Notices of the Royal Astronomical Society, 216, 17
\bibitem[5]{Bernardeau96} 
	Bernardeau, F., \& van de Weygaert, R.\ 1996, Monthly Notices of the Royal Astronomical Society , 279, 693 

\bibitem[6]{Bert87}
    Bertschinger ~E., 1987, The Astrophysical Journal, 323, 103
\bibitem[7]{Bremer04}
     P.-T. ~Bremer, H. ~Edelsbrunner, B. ~Hamann, and V. ~Pascucci. A topological hierarchy for functions on triangulated surfaces. IEEE Transactions on Visualization and 		Computer Graphics, 10(4):385Ð396, 2004
\bibitem[8]{Coles91} Coles, P., \& Jones, B.\ 1991, Monthly Notices of the Royal Astronomical Society , 248, 1 
\bibitem[9]{Delaunay34} Delaunay B., 1934, Bull. Acad. Sci. USSR (VII) Classe Sci. Mat., pp 793-800.

\bibitem[10]{Doi91}
	A. ~Doi, A. ~Koide, An Efficient Method of Triangulating Equivalued Surfaces by using Tetrahedral Cells.
	IEICE Transactions Communication, Elec. Info. Syst, E74(1) 214-224, January 1991    
\bibitem[11]{Edels03}
	H. ~Edelsbrunner, J. ~Harer, V. ~Natarajan and V. ~Pascucci. Morse-Smale complexes for piecewise linear 3-manifolds. 
	In Proc. 19th Ann. Sympos. Comput. Geom., pages 361Ð370, 2003
\bibitem[12]{got-mel-dic-86}
Gott III, J.R., Melott, A.L., Dickinson, M. 1986, The Astrophysical Journal, 306, 341
\bibitem[13]{Gueziec95}
	A. ~Gueziec, R. ~Hummel, Exploiting Triangulated Surface Extraction using Tetrahedral Decomposition.
	IEEE Transactions on Visualisation and Computer Graphics, 1 (4) 328-342, December 1995
\bibitem[14]{Hahn07a} 
	Hahn O., Porciani C., Carollo C.~M., Dekel A.\ 2007, Monthly Notices of the Royal Astronomical Society , 375, 489 
\bibitem[15]{jan-dem-83}
{Janes}, L.G. and {Demarque}, P., 1983, The Astrophysical Journal, 264, 206	
\bibitem[16]{Lorensen87} 
	W. ~Lorensen, H. Cline, Marching Cubes: A High Resolution 3D Surface Construction Algorithm.
     Computer Graphics, 21 (4): 163-169, July 1987
\bibitem[17]{lap-gel-huc-91}
	{de Lapparent}, V., {Geller}, M.J. \& {Huchra}, J.P., 1991, ApJ, 369, 273
        
\bibitem[18]{kandu}
	Malik, R. K. and Subramanian, K., 1997, Astronomy and Astrophysics, 317, 318.
\bibitem[19]{mat-03}
	Matsubara, T., 2003, ApJ, 584, 1
\bibitem[20]{mec-buc-wag-94} {Mecke}, K.R., {Buchert}, T. \&
 	 {Wagner}, H., 1994, Astronomy and Astrophysics 288, 697
    
\bibitem[21]{nov-etal-06}
	Novikov, D., Colombi, S. and Dor\'{e}, O., 2006, Monthly Notices of the Royal Astronomical Society, 366, 1201
\bibitem[22] {nov-fel-sh-99}
	{Novikov}, D.I., {Feldman}, H. and {Shandarin}, S.F., 1999 Int. J. Mod. Phys.D, 8, 291

\bibitem[23]{Okabe00}
	Okabe A., Boots B., Sugihara K., Chiu S. N., 2000, Spatial Tessellations, Concepts and Applications of Voronoi
	Diagrams. John Wiley, Chichester.

\bibitem[24]{pee-80} 
	{Peebles}, P.J.E. 1980, The Large-Scale Structure of the Universe(Princeton : Princeton Univ. Press)
\bibitem[25]{sah-sat-sh-98} {Sahni}, V., {Sathyaprakash}, B.S. \&
  	{Shandarin}, S.F., 1998, The Astrophysical Journal Letters, 495, L5
\bibitem[26]{sat-sah-sh-98} {Sathyaprakash}, B.S., {Sahni}, V. \&
  	{Shandarin}, S.F., 1998, The Astrophysical Journal, 508, 551
\bibitem[27]{Schaap00}
	Schaap W.~E., van de Weygaert R. 2000, Astronomy and Astrophysics, 363, L29
\bibitem[28]{Schaap07}
	Schaap, W.~E.\ 2007, The Delaunay Tessellation Field Estimator, Ph.D. Thesis, University of Groningen
\bibitem[29]{sch-buc-97} 
	{Schmalzing}, J.  \& {Buchert}, T., 1997, The Astrophysical Journal Letters, 482, L1
\bibitem[30] {sh-83} 
	Shandarin, S.F. 1983 Sov. Astron. Lett.,  9, 104
\bibitem[31]{sh-z-89} 
	{Shandarin}, S.F. and {Zeldovich}, Ya.B. 1989, Rev.Mod.Phys.,61, 185.
\bibitem[32]{sh-hab-hei-09}
  	Shandarin, S., Habib, S., Heitmann, K.  2009,   arXiv:0912.4471
\bibitem[33]{sh-etal-02}
	{Shandarin}, S.F., {Feldman}, H.A., Xu, Y., \& Tegmark, M. 2002, ApJS, 141, 1	
\bibitem[34]{sh-she-sah-04}
	Shandarin, S.F., Sheth, J., Sahni, V. 2004, Monthly Notices of the Royal Astronomical Society, 353, 162
\bibitem[35]{she-mn-04}
  	{Sheth}, J.V. 2004a, Monthly Notices of the Royal Astronomical Society, 354, 332
\bibitem[36]{she-phd-04}
 	 {Sheth}, J.V. 2004b, PhD Thesis, Pune University, Pune, India	
\bibitem[37]{she-etal-03}
  	{Sheth}, J.V., Sahni, V., Shandarin, S.F. \& Sathyaprakash, B. 2003, Monthly Notices of the Royal Astronomical Society, 343, 22
\bibitem[38]{Sibson81}	
	Sibson R., 1981, In: Barnet V.(ed.) Interpreting Multivariate Data. Wiley, Chichester
\bibitem[39]{sou-etal-08}
	Sousbie, T., Pichon, C., Colombi, S., Novikov, D.,  and Pogosyan, D.  2008, Monthly Notices of the Royal Astronomical Society, 383, 1655
\bibitem[40]{Springel05} 
  	Springel V., White S.D.M., Jenkins A., Frenk C.S., Yoshida N., Gao L., 
  	Navarro J., Thacker R., Croton D., Helly J., Peacock J.A., Cole S., 
  	Thomas P., Couchman H., Evrard A., Colberg J.M., Pearce F.\ 2005, Nature, 435, 629 
\bibitem[41]{Sugiyama95} 
	Sugiyama, N.\ 1995, \apjs, 100, 281
\bibitem[42]{Sukumar98}
	Sukumar N., 1998, PhD thesis, Northwestern University.
\bibitem[43]{Weygaert89} 
	van de Weygaert, R., Icke, V.\ 1989, \aap, 213, 1.
\bibitem[44]{Weygaert09}
      van de Weygaert R., Schaap W.E\ 2009, The Cosmic Web: Geometric Analysis.
      In: \textit{Data Analysis in Cosmology},  eds. V. Mart{\'{\i}}nez et al.,
      Lecture Notes Phys., 665, 291-413 (Springer-Verlag)
      arXiv:0708:1441	
\bibitem[45] {Voronoi1908} 
        Voronoi G., 1908, J. Reine Angew. Math., 134, 198

\bibitem[46]{Watson92} 
	Watson D. F., 1992, Contouring: A Guide to the Analysis and Display of Spatial Data. Pergamon Press.

\bibitem[47]{z-70}
	Zel'dovich, Ya.B. 1970, Astronomy and Astrophysics, 5, 84
\bibitem[48]{z-ein-sh-82}
	{Zel'dovich}, Ya.B., {Einasto}, J. and {Shandarin}, S.F., 1982, Nature, 300, 407

\end{thebibliography}
\end{document}